This is the author's version of the work. The definitive Version of Record was published in the *2025 IEEE International Conference on Systems, Man, and Cybernetics (SMC)*, Vienna, Austria, October 5–8, 2025. **DOI:** https://doi.org/10.1109/SMC58881.2025.11343025

# Designing Effective Human-Swarm Interaction Interfaces: Insights from a User Study on Task Performance

Wasura D. Wattearachchi[1,*], Erandi Lakshika[1], Kathryn Kasmarik[1], and Michael Barlow[1]
[1]University of New South Wales (UNSW) Canberra, School of Systems and Computing, Northcott Drive, ACT 2600, Australia.
[*]Corresponding author: Wasura D. Wattearachchi. E-mail: w.wattearachchi@unsw.edu.au

*Abstract*—**In this paper, we present a systematic method of design for human-swarm interaction interfaces, combining theoretical insights with empirical evaluation. We first derived ten design principles from existing literature, applying them to key information dimensions identified through goal-directed task analysis and developed a tablet-based interface for a target search task. We then conducted a user study with 31 participants where humans were required to guide a robotic swarm to a target in the presence of three types of hazards that pose a risk to the robots: Distributed, Moving, and Spreading. Performance was measured based on the proximity of the robots to the target and the number of deactivated robots at the end of the task. Results indicate that at least one robot was brought closer to the target in 98% of tasks, demonstrating the interface's success in fulfilling the primary objective of the task. Additionally, in nearly 67% of tasks, more than 50% of the robots reached the target. Moreover, particularly better performance was noted in moving hazards. Additionally, the interface appeared to help minimise robot deactivation, as evidenced by nearly 94% of tasks where participants managed to keep more than 50% of the robots active, ensuring that most of the swarm remained operational. However, its effectiveness varied across hazards, with robot deactivation being lowest in distributed hazard scenarios, suggesting that the interface provided the most support in these conditions.**

## I. Introduction

Effective Human-Swarm Interaction (HSI) interfaces are essential for enabling users to efficiently manage Multi-Robot Systems (MRS) in dynamic environments. However, designing an interface that maximises Situation Awareness (SA) and task performance remains a challenge. This paper presents a systematic approach to interface design for HSI, integrating theoretical foundations with empirical evaluation to develop and assess an interface tailored for a target search task.

In swarm robotics, cost-effective and often disposable robots are commonly deployed due to budget constraints [1]. However, operating such robots in dynamic environments with hazards increases the risk of deactivation. In these conditions, what ultimately matters is whether the user can still perform the task effectively using the available interface. While human-agent teaming has been explored in dynamic contexts, the specific interaction between humans and robotic swarms remains underexplored.

Although prior work has proposed design principles for HRI [2] and identified swarm-related considerations for HSI [3], many existing principles and taxonomies in MRSs emphasise what to design rather than how to design it. Addressing this gap, we introduced ten principles in this paper for effective HSI derived from the literature. Then, we focused on evaluating the effectiveness of a tablet-based HSI interface developed using those principles and the framework introduced in [4]. Effectiveness of the interface is assessed through users' task performance in a target search scenario, specifically by analysing how close robots were to the target at task completion, and how well users were able to minimise robot deactivations. Since the robots used are assumed to be expendable [5], successful task execution despite potential losses becomes a key indicator of interface effectiveness.

A large-scale study was conducted [4], examining both task performance and SA in detail. It was found that both subjective and objective SA were maintained across attempts, and certain SA elements (particularly Levels 1 and 3) were found to correlate with better task performance. However, as the analysis of SA is beyond the scope of this paper, this paper will solely focus on task performance. The main contributions of this paper are:

- A systematic approach to HSI interface design: We introduce ten principles derived from existing literature while integrating SA requirements into interface decisions, and then systematically designed a tablet-based user interface for HSI.
- An in-depth user study with human participants: Evaluation of task performance using the designed interface through a user study with 31 participants.

Our findings demonstrate how interface design impacts task performance, emphasising the need for adaptive support based on hazards. These insights are valuable for researchers developing HSI interfaces for complex, real-world applications. The rest of the paper is structured as follows: Section II covers background and related work, outlining principles for effective HSI. Section III details the interface design based on these principles. Section IV presents the user study, focusing on participants' task performance, and Section V concludes with key findings and future directions.

## II. Background and Related Work

This section examines key HSI techniques and existing design principles. It then explains how the principles proposed in this research were derived by integrating these insights, ensuring a more structured and effective approach to HSI design.

## A. Human-Swarm Interaction Techniques

HSI techniques offer various advantages and trade-offs depending on the approach. Those can be categorised based on how users influence swarm behaviour, communication methods, and observation strategies. Direct interaction methods, such as selecting swarm algorithms or manually controlling leaders [6], provide precise control and enable deliberate behavioural changes, but they can be cognitively demanding and may reduce the scalability and autonomy of the swarm. In contrast, indirect interaction methods—like environmental cues (e.g., virtual pheromones [7]), beacons [8], breadcrumbs [9] and geofencing [10] support more scalable and emergent behaviours with less operator input. However, these methods can be less intuitive and slower to produce desired outcomes due to their indirect nature. Communication mechanisms can be remote or proximal. Remote methods such as graphical user interfaces (GUIs) [6] and virtual reality [11] are well suited for hazardous environments, allowing operators to control swarms from a distance, but they may limit SA and suffer from latency. Proximal methods, including augmented reality, gestures, or voice commands [12], [13], offer more natural and immersive control, though they require the operator to be physically close and may be sensitive to environmental conditions. Finally, observation techniques, such as visualising individuals [14], subgroups [15], area-based perspectives [16] and heat-maps [17] and displaying summarised metrics like alignment and coverage aiding swarm assessment [18],[19], help reduce cognitive load and enhance decision-making. However, multi-screen setups with different access levels to monitor operations [10] will increase the cognitive load of the user if not carefully designed.

This study introduces a tablet interface with remote and mixed-method interaction, where users mark avoidance regions indirectly and guide robots directly with swipe gestures. A tablet interface was chosen for its portability, ease of use, and familiarity to search and rescue teams, making them suitable for outdoor tasks despite some limitations, such as sensitivity to sunlight, weather, and battery drain [10].

## B. Principles for Human-Swarm Interaction

Goodrich and Olsen [2] introduced a set of principles for designing MRSs, emphasising user-friendliness, efficient interaction, and reduced cognitive burden. These principles focus on enabling seamless mode transitions, leveraging natural signals, promoting direct manipulation of the environment rather than individual robots, and structuring information in a way that supports user attention and memory. Similarly, the research identified key features [3] for HSI, such as fostering decentralised intelligence, supporting scalability, enabling autonomous operation with human input, and designing interfaces that present relevant information to enhance situational awareness. While these principles and features define what HSI should achieve, they do not fully address how to design interactions that optimise performance. To bridge this gap, prior work has explored taxonomies for MRSs [20], categorising systems based on autonomy levels, human-to-robot ratios, interaction dynamics, decision-making support, and task complexity. However, these taxonomies primarily focus on system classification rather than guiding design decisions. A more effective approach would involve extending these taxonomies by addressing how swarm-specific characteristics can be integrated to enhance human-swarm teaming. Additionally, insights from Human-Computer Interaction (HCI) theories, such as Norman's seven stages of action [21], highlight how users interact with interfaces by setting goals, forming intentions, executing actions, and evaluating outcomes. Norman also identified key challenges in interface design, including mismatches between user expectations and system responses, which can create barriers to intuitive control. Prior work [22] has demonstrated the applicability of these principles in HRI, presenting an opportunity to incorporate them into HSI as well.

When discussing information, it's also important to consider the information gap—the mismatch between the large volume of data generated by complex systems and the operator's ability to access, interpret, and act on the most relevant information. In high-data environments like aviation or power systems, excessive and poorly organised information can overwhelm users, reducing rather than enhancing their SA [23]. Given SA's critical role in effective decision-making, our study addresses it in a separate manuscript that contains SA-related findings [4].

Building on these foundations, this work derives ten principles for efficient HSI. These provide a structured way to improve interaction effectiveness, ensuring that human operators can manage swarms more intuitively and effectively. A summary of these principles is presented in Table I.

TABLE I. TEN PRINCIPLES OF EFFICIENT HUMAN-SWARM INTERACTION

| Principle | Description |
|---|---|
| 1. Utilise local information and create a global picture out of it | Swarm systems are decentralised [3], relying on local information to form a global state. Interacting with a few members can influence the swarm's collective behaviour. |
| 2. Facilitate switching between swarm autonomy and human control | Scholtz [22] defined five HRI roles: supervisor, operator, teammate, bystander, and mechanic. In HSI, the supervisor and operator roles dominate as humans guide the swarm [3]. Full human control negates swarm autonomy, reducing it to an MRS, making a balance essential. Parasuraman et al. [24] outlined automation levels (1–10), which apply to HSI, allowing fine-tuned autonomy from full human control to full swarm independence, optimising collaboration. |
| 3. Use versatile interaction methods | Versatility in HSI includes scalability, flexibility, and extensibility. Scalability concerns factors like swarm size, operators, interfaces, area, and task complexity. Flexibility ensures interaction works for both homogeneous and heterogeneous swarms [25]. Extensibility aids operators in modifying goals, such as adding, changing, or removing them [25]. |
| 4. Tacitly switch between interaction modes, methods and tools | Goodrich and Olsen's model [2] includes "implicitly switches interfaces and autonomy modes". In our model, autonomy balance is separate (2nd principle), emphasising mode, method, and tool switching. This includes selecting swarming algorithms [6], adjusting sensory inputs [26], and shifting control methods like a GUI to a joystick [26]. |
| 5. Use intrinsic modalities over conventional | This principle promotes intrinsic modalities over traditional interfaces, using natural inputs like voice [27], gestures [12], and touch [9] for |
| methods and interfaces | intuitive swarm interaction. Goodrich and Olsen [2] also emphasised naturalistic communication. Trust calibration cues—audio, visual, anthropomorphic, and verbal—enhance trust in swarming systems [28]. Immersive interfaces like VR [29] and AR [30] enhance interaction by merging digital and physical worlds. Haptic interfaces provide tactile feedback, aiding swarm behaviour learning [31]. Multimodal interfaces combine inputs like voice, gestures, VR, AR, and touch for flexible control [27]. Moreover, tablet interfaces [10] that facilitate portability and higher familiarity with search and rescue teams are useful in this context. |
| 6. Minimise the interaction gap between the swarm member(s) and the human during direct interaction | Yanco and Drury [20] explored HRI types, but their taxonomy omitted direct vs. indirect methods. Since swarms are MRSs, key interactions include one human–many robots, human team–robot team, and many humans–robot team. Direct interaction methods help influence swarms, and this principle aims to simplify HSI. When selecting direct techniques, factors like robot morphology (anthropomorphic, zoomorphic, or functional) and swarm homogeneity or heterogeneity [20] should also be considered. |
| 7. Minimise the interaction gap between the environment and the human during indirect interaction | Indirect interaction also enables human influence on the swarm. Yanco and Drury [20] did not focus on this, so this principle focuses on reducing interaction complexity between humans and the environment. As with direct interaction, swarm composition [20] (homogeneous or heterogeneous) should be considered when selecting indirect methods. |
| 8. Enable engagement with the information | Goodrich and Olsen's Model [2] emphasises interactive interfaces in HRI. Instead of static displays, interfaces should allow interaction—e.g., clicking a map element should adjust the user's view accordingly. |
| within the interface | |
| 9. Reduce what should be remembered | Goodrich and Olsen's Model [2] also highlights externalising memory. Users shouldn't rely solely on recall; past data and decisions should be recorded for reference when needed. |
| 10. Assistance in managing the attention | Attention management is crucial in HSI, as noted in Goodrich and Olsen's Model [2]. Given swarm complexity, user attention can be diverted or tunnelled [26]. Interfaces should highlight key information to aid SA. |

## III. Interface Design

A tablet-based user interface was developed to facilitate human interaction with a swarm of robots in a simulation, with the design rooted in SA principles. The focus of this interface was to support a swarming task, which involves scenarios such as target search, navigation, or adversarial patrolling. For this study, a task was designed where 20 robots search for a single target using the PSO LBEST model [32], with the environment simulated in CoppeliaSim [33]. Participants interacted with this environment via a tablet interface, as shown in Figure 1. Humans received hazard alerts via the tablet and guided robots to avoid these areas to prevent their deactivation. After defining the task, an SA requirement analysis was carried out using GDTA, by identifying Goals, Sub-goals, Decisions and SA requirements. Then those were abstracted into information dimensions based on the type of knowledge that they present. The detailed process is included in our manuscript, which contains a detailed analysis of SA [4].

Table II presents the information dimensions (Dim1: Location information of robots and the target, Dim2: Motion and spatial state information of the robots, Dim3: Temporal progress information, Dim4: Dynamic obstacles status information, Dim5: Robot loss information, Dim6: Robot trapped information) derived using the standard GDTA and outlines how the design principles have been applied to each dimension to decide the interface elements and interactions. It is important to note that not all principles are necessarily applied to every information dimension. In the table, Pi denotes the application of the i-th principle to a specific information dimension.

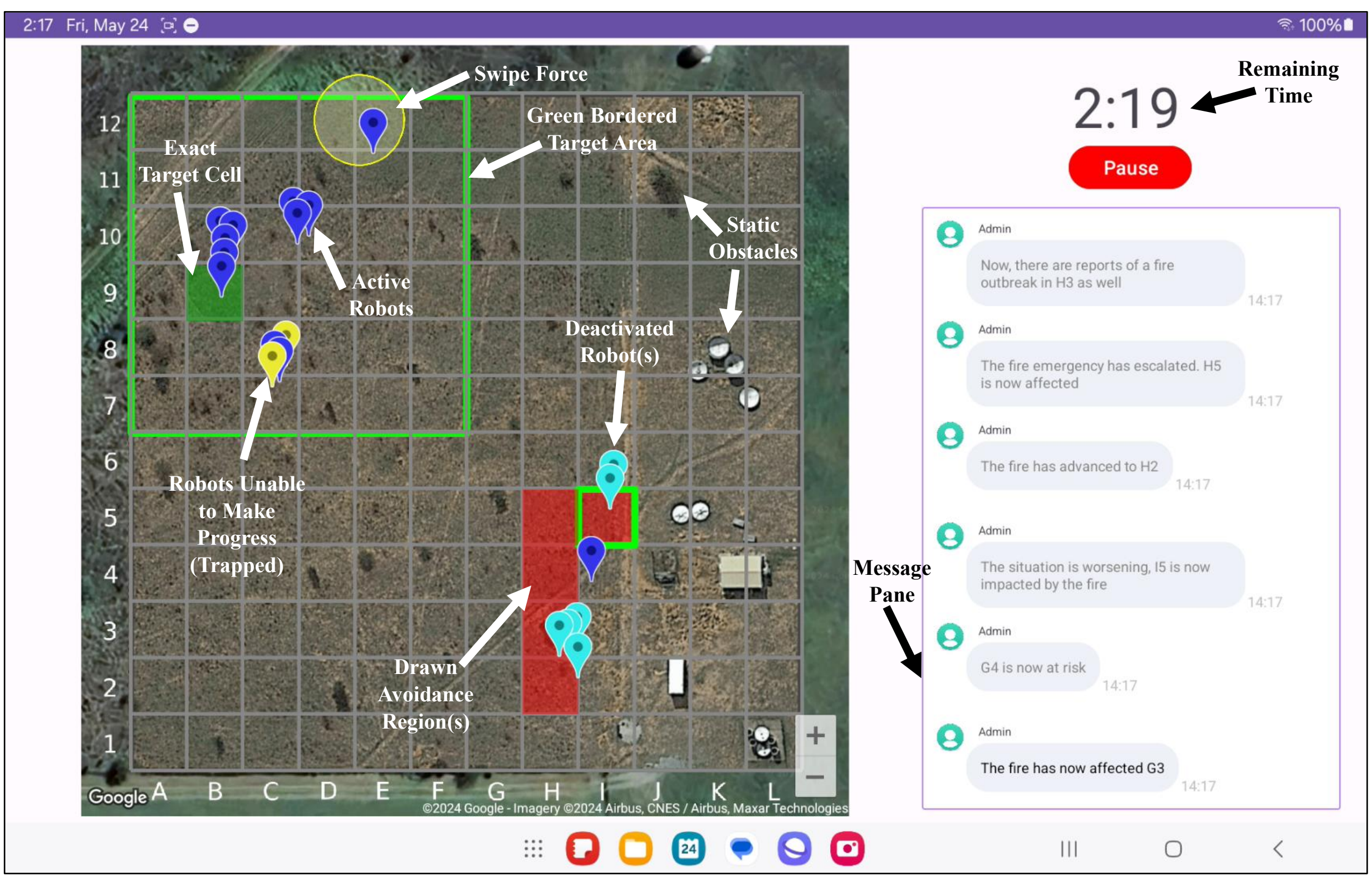

Figure 1. Designed tablet interface [4]

TABLE II. DESIGNING INTERACTIONS FOR THE TARGET FINDING SWARMING TASK FOR EACH INFORMATION DIMENSION

| Dim | Design Decision |
|---|---|
| *Dim1* | P1: Robots use PSO LBEST, sharing local best positions within a set radius, leading to emergent global behaviour.<br>P3: The interface will update automatically to display new swarm members if added.<br>P6: No direct interaction, since it is difficult to click and select one robot when the robots are close to each other.<br>P8: Zoom in/out via gestures or icons for better visibility.<br>P9: Robots and obstacles are visualised, eliminating the need for memory recall; the target area is highlighted.<br>P10: Zoom in to focus on a specific set of robots. |
| *Dim2* | P1: Robots use PSO LBEST, relying on local data.<br>P2: The operator can influence robot motion by marking hazardous areas without breaking autonomy.<br>P7: Interaction lag in marking regions is minimised.<br>P9: Robot movement updates in real-time on the map.<br>P10: Zoom in to track specific moving robots. |
| *Dim3* | P1: Underline PSO uses individual robots' progress and state (for example, local best position in a neighbourhood) with time when deciding the movement direction and the velocity of the other neighbouring robots.<br>P2: Swarm autonomy and human interaction will happen simultaneously (For instance, robots continue their task while the operator interacts).<br>P3: Task time extensions must be communicated to the operator via the timer.<br>P9: Remaining time is displayed to avoid memory recall.<br>P10: Time display colour changes as the task nears completion to get the user's attention. |
| *Dim4* | P1: Each robot has obstacle avoidance and a minimum sensing distance, enabling the entire swarm to avoid dynamic obstacles.<br>P2: The operator can mark cells so that the robots will avoid those regions using their autonomy.<br>P3: Multiple hazard zones can be available at one instance and can be sequentially marked by the operator. The message pane that shows alerts can contain more than one message.<br>P5: Single tap to mark a hazardous cell, long press to delete, which will prompt a confirmation dialogue asking whether the user needs to delete the region or not.<br>P7: Minimise lag when marking/deleting hazard cells. The marked cells will act as obstacles to robots, so they will move away from those, which denotes an indirect effect on the swarm.<br>P8: Tapping a marked cell will highlight it; long press prompts deletion confirmation to show engagement.<br>P9: Marked areas are mapped to the search space; notifications appear in chronological order.<br>P10: Notifications include grid coordinates for precise marking and narrow down user attention. When a user taps on a marked cell, its border colour will change, showing that it is currently selected. Users can zoom in on a particular area of interest and mark cells if they want. |
| *Dim5* | P1: Deactivated robots stop contributing to PSO, altering swarm behaviour.<br>P9: Deactivated robots are displayed in light blue.<br>P10: Light blue deactivated robots draw attention; zooming allows focus on affected areas. |
| *Dim6* | P2: Swiping moves nearby robots briefly, allowing the robots in the direction of the swipe for a few seconds, which symbolises the switching between autonomy and human control.<br>P3: Multiple swipes apply cumulative force to the neighbouring robots.<br>P6: Swipe force stops when the user lifts their finger.<br>P7: Swipe affects all robots within a set radius.<br>P9: Trapped robots appear yellow for quick identification.<br>P10: When swipe forces are applied, those will be shown in yellow circles. |

When discussing the interface elements and the interactions, the robots are initially active, denoted in dark blue as shown in Figure 1, and they try to find the target (shown as the dark green cell), which will be only revealed to the user when at least one of the robots reaches it. Until then, they will be shown the roughly marked target area with the green border. Users will be informed about the hazards via messages, and if robots enter those areas, they will be deactivated (dead, shown in light blue). To avoid deactivation, users can tap on specific cells to mark them in red, so that the robots will avoid them. Yellow robots are trapped robots, which can be activated again by swiping near them, which will apply a force to all neighbouring robots within a defined radius (including active robots).

## IV. USER STUDY AND RESULTS

With the developed interface, a user study was conducted (with the approval of the Human Research Ethics Approval Panel at UNSW Canberra) involving 31 participants (15 males, 15 females, 1 non-disclosing) aged 18 to 44, primarily postgraduate and undergraduate students. Most (25) were aged 25 to 34, with 27 non-native English speakers. The participants' knowledge of robotic swarms followed a normal distribution. Participants used the tablet interface to control RS in three scenarios: distributed, moving, and spreading hazards. In the spreading scenario, hazards expanded from a single point, such as a spreading fire, while in the distributed scenario, they appeared intermittently, such as falling rocks and trees, loss of GPS signals, strong winds, or damaged terrain. Moving hazards are regions that shift throughout the task, such as adversarial patrolling. The primary variation among these scenarios lies in the type of alert messages displayed on the screen.

First, they completed a pre-task questionnaire via Qualtrics [34], providing demographic details (age, gender, education, touch device familiarity, and knowledge of RS). Then participants completed three five-minute tasks for each hazard type, with two pauses per task to answer 14 SAGAT [35] questions measuring objective SA across Endley's three SA levels in each pause (during one task), altogether 28 SAGAT questions. Questions were multiple choice or cell-marking, with marks ranging from 0 to 100; partial marks were given for proximity in cell-marking questions.

After each task, participants completed a subjective SA assessment using the 10D SART [36], rating 10 questions on a 7-point Likert scale. Tasks were repeated in a second attempt in a randomised order, following the same procedures. This process resulted in each participant performing a total of six tasks, referred to as "participant-tasks (PT)". The study took approximately two hours per participant. All results that involved significance testing were reported at a 95% confidence level, with $p < 0.05$ considered significant.

### A. Experiment 1: Examining the task performance based on proximity to the target

This section discusses the task performance of the participants first when considering all the tasks together, then considering each hazard separately.

#### 1) Analysis of performance considering all hazards:

**Aim:** To determine how effective the interface is in target localisation by examining various task performance metrics based on the proximity of active robots to the target by the end of the task.

**Hypothesis:** The user interface will be effective in target localisation, as indicated by a higher proportion of participant tasks in which active robots are positioned within 0 to 12m of the target across different task performance metrics. This includes the position of the nearest active robot (N) to the target at the end of the task, as well as the 25th (NQ1), 50th (NQ2), and 75th (NQ3) percentile distances of all active robots, providing insight into how well different portions of the swarm were guided. It also includes the farthest active robot (F) and the centroid of all active robots (C), capturing overall swarm dispersion and control.

**Method:** Task performance was measured based on how close the positions of active robots were to the target at the end of the task. The N, NQ1, NQ2 and NQ3 reflect the user's ability to accurately and precisely guide robots close to the target. F indicates the user's ability to manage the entire group of robots, ensuring that even the most distant one is relatively close to the target. Meanwhile, C represents the user's ability to guide the active robots cohesively towards the target. Since these scores reflect the distance from the target, a higher score indicates a lower task performance and vice versa. All PT were analysed together by examining their task performance. We then assessed how many PT participants brought robots closer to the target based on each task performance metric. The 0 to 12 m range was chosen as the closest, as it represents 10% of the arena's width. Exploratory checks with nearby thresholds (e.g., 10 m and 15 m) did not materially change the overall pattern of findings.

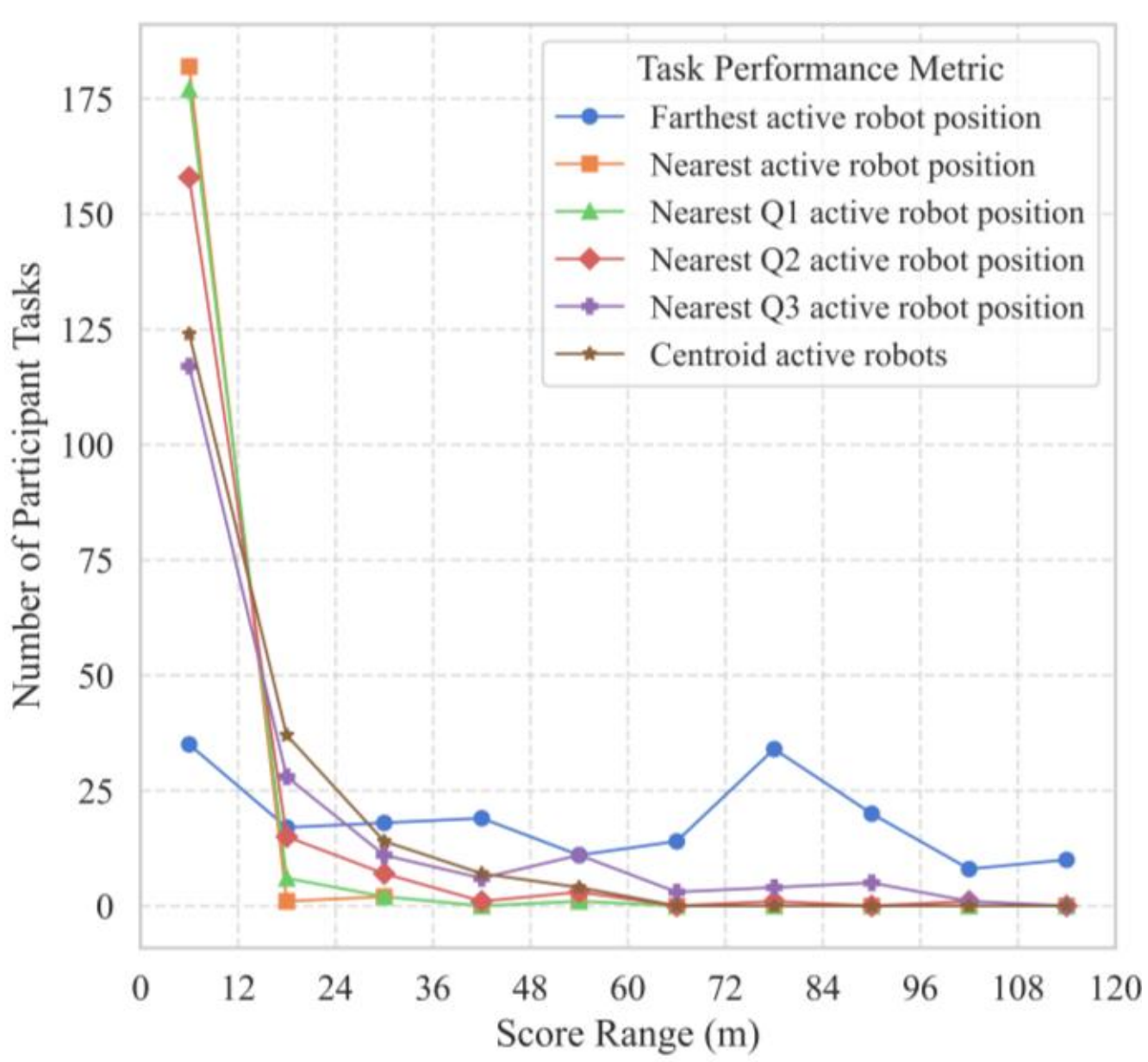

Figure 2. Task performance metrics are based on how close the robots are to the target. The score range is from 0 to 120, representing the distance between the target and the robots within a 120x120 space. Lower distances indicate better performance, with 0 being best and 120 being the worst.

**Results:** Figure 2 shows the distribution of PT scores (186 PT) based on the task performance metrics (shown as histograms where only the centre points of each bin are connected with lines).

Based on this, the results indicate that the user interface effectively assisted participants in guiding at least one robot close to the target, as evidenced by 97.85% of PT achieving this for the N. In addition, the results for C show that 66.67% of PT were able to bring C within the 0 to 12 m range. Since this score is above 50%, it suggests that more than half of the PT could guide the overall group of active robots relatively close to the target. This serves as a baseline indicating a moderate level of effectiveness in managing the swarm collectively. However, when examining the participants' ability to guide the NQ1, NQ2 and NQ3, a gradual decline in performance is noticeable. Specifically, 95.16% of PT managed to bring the NQ1 within the 0 to 12 m range, which drops to 84.95% for the NQ2 and further decreases to 62.90% for the NQ3. This trend suggests that while users could bring some robots close to the target, consistently bringing them closer to the target across all quartiles diminished. The results for F reinforce this observation, with only 18.82% of tasks achieving the 0 to 12 m range for the most distant robots. This indicates that the user interface was less effective in assisting users to guide the entire swarm close to the target.

**Conclusion:** The results support the hypothesis that the user interface was effective in target localisation, as evidenced by a higher proportion of participant tasks in which active robots were positioned within the closest range (0 to 12m) of the target. The findings indicate that at least one robot reached this range in 98% of tasks. Additionally, 50% of the robots (NQ2) were brought within this range in 85% of tasks, suggesting that users could effectively guide a significant portion of the swarm closer to the target. However, participants were unable to bring all the robots always. Considering the nature of a target search task, the primary objective was to ensure that at least some robots reached the target. In this regard, the user interface succeeded in providing the necessary support for users to achieve this goal.

#### 2) Analysis of performance by hazard:

**Aim:** To assess how effectively the user interface assisted participants in guiding a swarm of robots towards a target across different hazards.

**Hypothesis:** The user interface is more effective in certain hazard types when considering target localisation, as indicated by significantly higher task performance in some hazards compared to others.

**Method:** To assess task performance across different hazards, we analysed six distance-based metrics similar to Experiment 1.1. For statistical analysis, we first computed the median values for each metric per participant across the two attempts and then got an average as shown in Table III. Since each participant completed the same hazard twice, we used the user identifier to pair their results before conducting comparisons. A Friedman test was performed to assess overall differences across hazards, followed by post-hoc pairwise comparisons (Wilcoxon signed-rank test) to identify significant differences, as shown in Table IV.

**Results:** Across the three hazards, Spreading consistently showed higher distances compared to Moving and Distributed, confirming that participants did not perform well in bringing robots closer to the target in this condition. For the N, Spreading had a median distance of 0.45 meters (IQR: 0.31–0.81), which was proven significantly higher than Distributed (0.32 meters, IQR: 0.14–0.45) and Moving (0.34 meters, IQR: 0.23–0.55). F also demonstrated this trend, with Spreading having a median distance of 74.32 meters (IQR: 55.96–83.88), while Distributed (46.09 meters, IQR: 33.89–57.81) and Moving (42.08 meters, IQR: 23.85–55.85) showed significantly lower distances. Similarly, C had a median distance of 16.67 (IQR: 11.43–21.76) in Spreading, which was significantly higher than Distributed (5.80 meters, IQR: 3.69–9.13) and Moving (5.95 meters, IQR: 4.54–11.20). These results indicate that participants performed worse (as indicated by higher distances) in the Spreading task, as robots were farther from the target in all key distance metrics.

**Conclusion:** The analysis supports the claim that the user interface was most effective in the Moving and Distributed tasks, while the Spreading task posed the greatest challenge for participants. Participants kept robots significantly closer to the target in the Moving and Distributed tasks compared to Spreading, as indicated by multiple significant differences in N, C, and F. The lack of significant differences between the Moving and Distributed tasks suggests that both were equally well supported by the interface. Future improvements could focus on enhancing control mechanisms for tasks with spread-out robots to improve task performance.

TABLE III. PROXIMITY TO THE TARGET FOR EACH HAZARD IN METERS - MEDIANS AND INTERQUARTILE RANGES (IQRS)

| Metric | Distributed | Moving | Spreading |
|---|---|---|---|
| **N** | 0.32 (0.14, 0.45) | 0.34 (0.23, 0.55) | 0.45 (0.31, 0.81) |
| **NQ1** | 1.39 (1.21, 1.88) | 1.38 (1.29, 1.90) | 1.70 (1.45, 4.74) |
| **NQ2** | 2.48 (2.17, 4.09) | 2.30 (2.08, 3.70) | 7.19 (2.61, 11.48) |
| **NQ3** | 5.71 (3.51, 11.60) | 7.70 (2.85, 23.65) | 24.57 (11.50, 34.24) |
| **F** | 46.09 (33.89, 57.81) | 42.08 (23.85, 55.85) | 74.32 (55.96, 83.88) |
| **C** | 5.80 (3.69, 9.13) | 5.95 (4.54, 11.20) | 16.67 (11.43, 21.76) |

TABLE IV. DIFFERENCE OF PROXIMITY TO THE TARGET BETWEEN HAZARDS IN FRIEDMAN TESTS AND POST-HOC PAIR-WISE COMPARISONS (DF = 2)

| Metric | Friedman Test Result (p-value < 0.05) | Post-Hoc Significant Differences (p < 0.05) |
|---|---|---|
| **N** | **p = 0.00** | Spreading > Distributed (p = 0.01), Spreading > Moving (p = 0.02) |
| **NQ1** | **p = 0.03** | Spreading > Distributed (p = 0.02) |
| **NQ2** | **p = 0.00** | Spreading > Moving (p = 0.02) |
| **NQ3** | **p = 0.02** | Spreading > Distributed (p = 0.00), Spreading > Moving (p = 0.02) |
| **F** | **p < 0.0001** | Spreading > Distributed (p = 0.00), Spreading > Moving (p < 0.0001) |
| **C** | **p = 0.00** | Spreading > Distributed (p = 0.00), Spreading > Moving (p = 0.00) |

*B. Experiment 2: Examining the task performance of the participants based on how many robots got deactivated*

This section discusses the task performance of the participants first when considering all the tasks together, then considering each hazard separately.

*1) Analysis of performance considering all hazards:*

**Aim:** To determine how the interface effectively helps the user in minimising robot deactivations by examining various task performance metrics based on assessing the percentage of robots deactivated during PT.

**Hypothesis:** The user interface is effective in minimising robot deactivations, ensuring that the majority of the swarm remains active throughout the task.

**Method:** Out of the 20 total number of robots, the number of deactivated robots per task was recorded. The percentage of tasks falling into different robot loss categories was calculated to assess how frequently participants maintained a high number of active robots.

**Results:** As shown in Figure 3, 7.0% of tasks had no robot losses, while 44.6% had minimal losses (1-5 deactivated). Moderate losses (6-10 deactivated) occurred in 42.5% of tasks, and higher losses (11-15 deactivated) were rare (5.9%). No extreme failures (16+ deactivated) occurred, suggesting the interface managed to keep at least one robot active to do the task.

**Conclusion:** The results continue to support the hypothesis that the interface helped participants minimise robot deactivation by keeping at least one robot active to do the task.

*1) Analysis of performance by hazard:*

**Aim:** To determine whether the interface assistance effectively minimises the number of deactivated robots during different hazards. By analysing the distribution of deactivated robots, we aim to identify which hazard benefits the most from interface assistance and which experiences the highest number of deactivations.

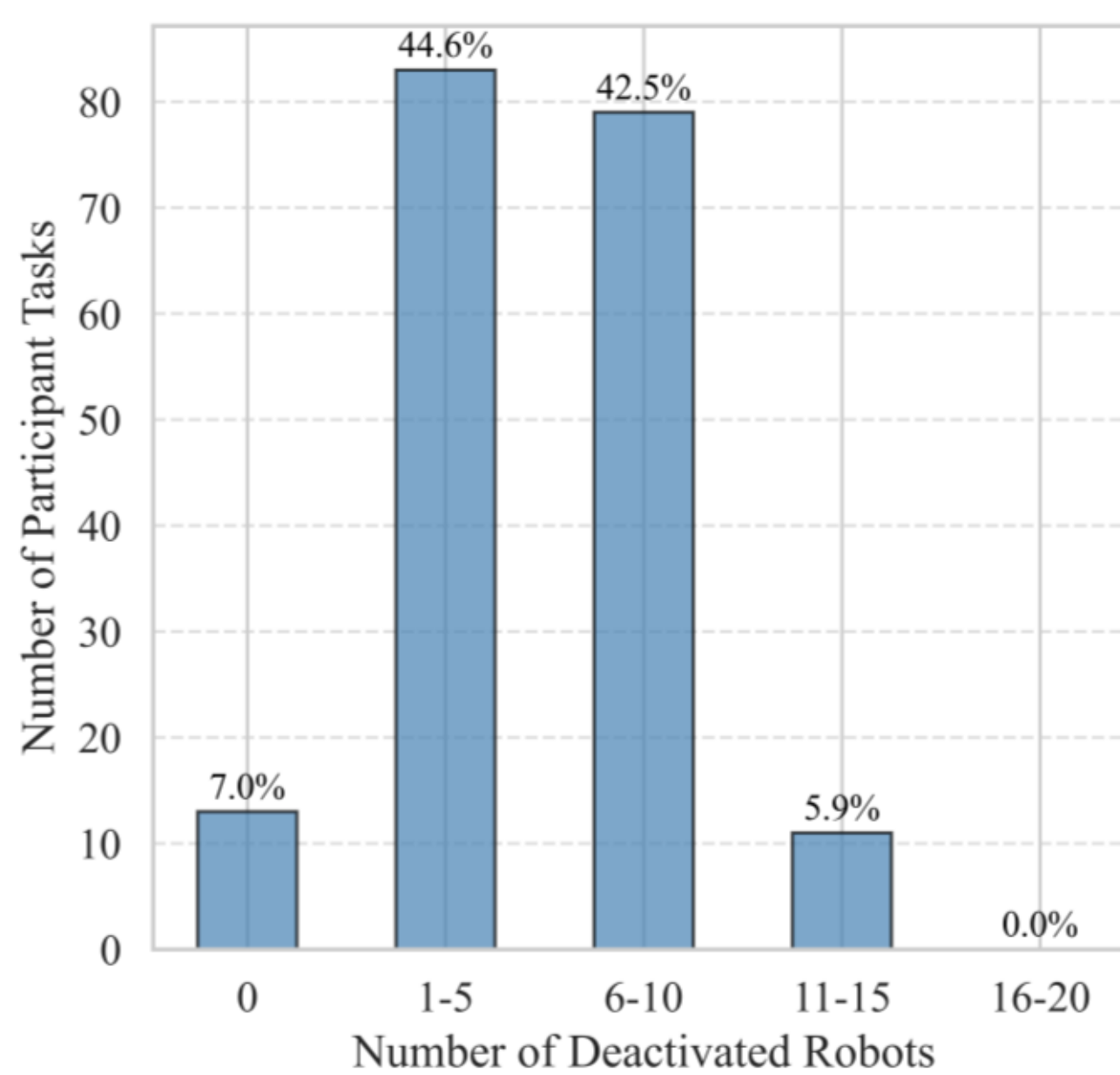

Figure 3. Distribution of deactivated robots across participant tasks. The histogram categorises tasks based on the number of robots deactivated, grouped into five bins: 0, 1-5, 6-10, 11-15, and 16-20. The percentage above each bar represents the proportion of tasks falling within each range.

**Hypothesis:** The user interface is more effective in certain hazard types in minimising robot deactivations, as indicated by significantly higher task performance in some hazards compared to others

**Method:** Similar to Experiment 1.2, but instead of distance-based metrics, we analysed the number of deactivated robots across the three hazards.

**Results:** The distribution of deactivated robots across the three hazards is shown in Figure 4. Each hazard has 62 occurrences, and the number of tasks falling into each deactivation category is shown as both counts and percentages.

The Moving task had a median of 7.00 deactivated robots (IQR: 6.50–8.00), which was significantly higher than both Spreading (4.50, IQR: 4.00–6.25) and Distributed (3.50, IQR: 2.50–5.00), which was confirmed by post-hoc pairwise comparisons as shown in Table V. Also, Spreading had significantly more deactivations than Distributed, reinforcing that task complexity influenced robot loss. These findings suggest that the interface was most helpful in Distributed tasks, where deactivations remained lowest, but provided less support in Moving tasks, leading to the highest failure rates.

**Conclusion:** The results indicate that the interface did not minimise deactivations equally across all hazards. Instead, Distributed had significantly fewer deactivations compared to both Moving and Spreading, suggesting that the interface was most effective in structured environments. In contrast, Moving experienced the highest deactivation rates, indicating that the interface provided less support in high-mobility scenarios. These findings suggest that the interface's effectiveness varies by hazard, with room for improvement in dynamic environments.

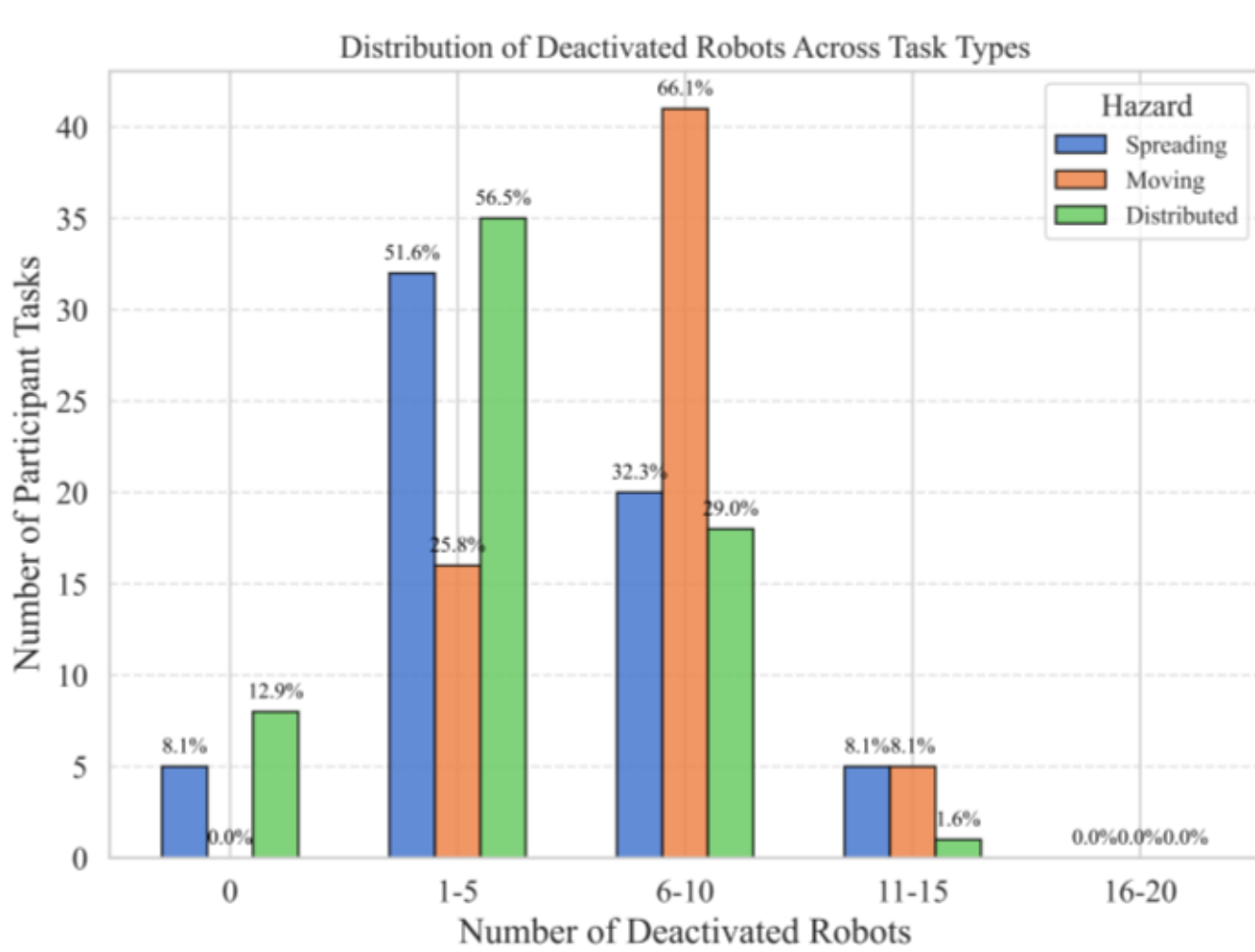

Figure 4. Distribution of deactivated robots for each task. The histogram groups the number of deactivated robots into five bins (0, 1–5, 6–10, 11–15, and 16–20) for each task. Each bin contains three bars representing the counts for different hazards, with percentages shown above.

TABLE V. DIFFERENCE IN NUMBER OF ROBOT DEACTIVATIONS BETWEEN HAZARDS IN POST-HOC PAIR-WISE COMPARISONS

| Comparison | p-value < 0.05 |
|---|---|
| **Distributed vs. Moving** | **< 0.0001** |
| **Distributed vs. Spreading** | **0.01** |
| **Moving vs. Spreading** | **0.00** |

## V. CONCLUSION, LIMITATIONS AND FUTURE WORK

This paper presents a systematic approach to designing and evaluating an interface for HSI. We first derived ten design principles based on existing theories in HRI, MRSs, and HSI. These principles were then systematically applied to the information dimensions identified using GDTA, leading to well-informed design decisions for a tablet-based interface.

To assess the interface's effectiveness, we conducted a user study with 31 participants performing a target search task in the presence of three hazard types: Distributed, Moving, and Spreading. Task performance was evaluated based on how close the robots were to the target and how many robots were deactivated by the end of the task.

The results show that the interface successfully assisted participants in guiding robots to the target, with at least one robot reaching the closest range (0–12 m) in nearly 98% of tasks. Additionally, nearly 67% of tasks resulted in more than 50% of robots reaching this range, indicating a moderate level of effectiveness in guiding the swarm collectively. The interface was particularly effective in scenarios with moving hazards, where better performance was observed. Furthermore, it helped minimise robot deactivation, as nearly 94% of tasks maintained more than 50% of robots active, ensuring that most of the swarm remained operational. Robot deactivation was lowest in scenarios with distributed hazards, suggesting that the interface provided the most support in these conditions. These findings highlight the need to adapt interface support to different task types to optimise performance.

This work provides a structured design framework for HSI interfaces, offering a systematic approach to integrating SA requirements into interface decisions. The empirical evaluation contributes actionable insights for researchers and designers developing interfaces for HSI, particularly in dynamic environments.

This study was conducted in a simulated environment and primarily involved Higher Degree Research students who were not experts in robot teleoperation. Their behaviour may differ from that of expert users, and they may adapt differently over time. Real-world factors such as hardware latency, sensor noise, and physical robot failures were not accounted for. Future work could address these limitations by testing the interface in physical environments and involving a broader participant pool. Additionally, efforts will focus on refining the interface for more complex scenarios such as Spreading and Moving hazards, incorporating adaptive features that respond to user workload in real time, and exploring shared control strategies and long-term user adaptation through longitudinal studies.

## REFERENCES

[1] S. G, A. S. A, and B. H. V, "A Survey on Swarm Robotic Modeling, Analysis and Hardware Architecture," *Procedia Computer Science,* vol. 133, pp. 478-485, 2018/01/01/ 2018.

[2] M. A. Goodrich and D. R. Olsen, "Seven principles of efficient human robot interaction," in *SMC'03 Conference Proceedings. 2003 IEEE International Conference on Systems, Man and Cybernetics. Conference Theme - System Security and Assurance (Cat. No.03CH37483)*, 8-8 Oct. 2003 2003, vol. 4, pp. 3942-3948 vol.4.

[3] S. Bashyal and G. K. Venayagamoorthy, "Human swarm interaction for radiation source search and localization," in *2008 IEEE Swarm Intelligence Symposium*, 21-23 Sept. 2008 2008, pp. 1-8.

[4] W. D. Wattearachchi, E. Lakshika, K. Kasmarik, and M. Barlow, "A Study on Human-Swarm Interaction: A Framework for Assessing Situation Awareness and Task Performance," *arXiv preprint arXiv:2503.14810,* 2025.

[5] A. Eudes, S. Bertrand, J. Marzat, and I. Sarras, "Distributed Control for Multi-Robot Interactive Swarming Using Voronoi Partioning," *Drones,* vol. 7, no. 10, p. 598, 2023.

[6] G. Kapellmann-Zafra, N. Salomons, A. Kolling, and R. Groß, "Human-robot swarm interaction with limited situational awareness," in *Swarm Intelligence: 10th International Conference, ANTS 2016, Brussels, Belgium, September 7-9, 2016, Proceedings 10*, 2016: Springer, pp. 125-136.

[7] M. Daily, C. Youngkwan, K. Martin, and D. Payton, "World embedded interfaces for human-robot interaction," in *36th Annual Hawaii International Conference on System Sciences, 2003. Proceedings of the*, 6-9 Jan. 2003 2003, p. 6 pp.

[8] J. McLurkin, J. Smith, J. Frankel, D. Sotkowitz, D. Blau, and B. Schmidt, "Speaking Swarmish: Human-Robot Interface Design for Large Swarms of Autonomous Mobile Robots," in *AAAI spring symposium: to boldly go where no human-robot team has gone before*, 2006: Palo Alto, CA, pp. 72-75.

[9] S. J. Bowley and K. Merrick, "A 'Breadcrumbs' Model for Controlling an Intrinsically Motivated Swarm Using a Handheld Device," Cham, D. Alahakoon and X. Li, Eds., 2017: Springer International Publishing, in AI 2017: Advances in Artificial Intelligence, pp. 157-168.

[10] M.-T. O. Hoang, N. v. Berkel, M. B. Skov, and T. R. Merritt, "Challenges and Requirements in Multi-Drone Interfaces," presented at the Extended Abstracts of the 2023 CHI Conference on Human Factors in Computing Systems, Hamburg, Germany, 2023.

[11] P. Walker, J. Hamell, C. A. Miller, J. Ladwig, H. Wauck, and P. K. Keller, "Immersive Interaction Interface (I3): A Virtual Reality Swarm Control Interface," *IEEE Transactions on Field Robotics,* vol. 1, pp. 424-445, 2024.

[12] A. Suresh and S. Martínez, "Gesture based Human-Swarm Interactions for Formation Control using interpreters," *IFAC-PapersOnLine,* vol. 51, no. 34, pp. 83-88, 2019/01/01/ 2019.

[13] S. Pourmehr, V. M. Monajjemi, R. Vaughan, and G. Mori, ""You two! Take off!": Creating, modifying and commanding groups of robots using face engagement and indirect speech in voice commands," in *2013 IEEE/RSJ International Conference on Intelligent Robots and Systems*, 3-7 Nov. 2013 2013, pp. 137-142.

[14] S. Nunnally *et al.*, "Human influence of robotic swarms with bandwidth and localization issues," in *2012 IEEE International Conference on Systems, Man, and Cybernetics (SMC)*, 14-17 Oct. 2012 2012, pp. 333-338.

[15] R. Fernandez Rojas *et al.*, "Encephalographic Assessment of Situation Awareness in Teleoperation of Human-Swarm Teaming," Cham, K. W. Wong and M. Lee, Eds., 2019: Springer International Publishing, in Neural Information Processing, pp. 530-539.

[16] Y. Lim *et al.*, "Avionics Human-Machine Interfaces and Interactions for Manned and Unmanned Aircraft," *Progress in Aerospace Sciences,* vol. 102, pp. 1-46, 2018/10/01/ 2018.

[17] M. Divband Soorati, J. Clark, J. Ghofrani, D. Tarapore, and S. D. Ramchurn, "Designing a User-Centered Interaction Interface for Human–Swarm Teaming," *Drones,* vol. 5, no. 4, p. 131, 2021.

[18] L. G. Jeston-Fenton, S. Abpeikar, and K. Kasmarik, "Visualisation of Swarm Metrics on a Handheld Device for Human-Swarm Interaction," in *2022 26th International Conference Information Visualisation (IV)*, 19-22 July 2022 2022, pp. 149-154.

[19] C. E. Harriott, A. E. Seiffert, S. T. Hayes, and J. A. Adams, "Biologically-Inspired Human-Swarm Interaction Metrics," *Proceedings of the Human Factors and Ergonomics Society Annual Meeting,* vol. 58, no. 1, pp. 1471-1475, 2014.

[20] H. A. Yanco and J. Drury, "Classifying human-robot interaction: an updated taxonomy," in *2004 IEEE international conference on systems, man and cybernetics (IEEE Cat. No. 04CH37583)*, 2004, vol. 3: IEEE, pp. 2841-2846.

[21] D. A. Norman and S. W. Draper, *User Centered System Design; New Perspectives on Human-Computer Interaction*. L. Erlbaum Associates Inc., 1986.

[22] J. Scholtz, "Theory and evaluation of human robot interactions," in *36th Annual Hawaii International Conference on System Sciences, 2003. Proceedings of the*, 6-9 Jan. 2003 2003, p. 10 pp.

[23] M. R. Endsley and D. J. Garland, *Situation awareness analysis and measurement*. CRC press, 2000.

[24] R. Parasuraman, T. B. Sheridan, and C. D. Wickens, "A model for types and levels of human interaction with automation," *IEEE Transactions on Systems, Man, and Cybernetics - Part A: Systems and Humans,* vol. 30, no. 3, pp. 286-297, 2000.

[25] C. Vasile, A. Pavel, and C. Buiu, "Integrating human swarm interaction in a distributed robotic control system," in *2011 IEEE International Conference on Automation Science and Engineering*, 24-27 Aug. 2011 2011, pp. 743-748.

[26] J. M. Riley, L. D. Strater, S. L. Chappell, E. S. Connors, and M. R. Endsley, "Situation awareness in human-robot interaction: Challenges and user interface requirements," *Human-Robot Interactions in Future Military Operations,* pp. 171-192, 2010.

[27] B. Larochelle, G. J. M. Kruijff, N. Smets, T. Mioch, and P. Groenewegen, "Establishing human situation awareness using a multi-modal operator control unit in an urban search & rescue human-robot team," in *2011 RO-MAN*, 31 July-3 Aug. 2011 2011, pp. 229-234.

[28] K. Okamura and S. Yamada, "Adaptive Trust Calibration for Supervised Autonomous Vehicles," presented at the Adjunct Proceedings of the 10th International Conference on Automotive User Interfaces and Interactive Vehicular Applications, Toronto, ON, Canada, 2018.

[29] A. Bhattacharya and S. Butail, "Designing a Virtual Reality Testbed for Direct Human-Swarm Interaction in Aquatic Species Monitoring," *IFAC-PapersOnLine,* vol. 55, no. 37, pp. 295-301, 2022/01/01/ 2022.

[30] V. Costa, M. Duarte, T. Rodrigues, S. M. Oliveira, and A. L. Christensen, "Design and development of an inexpensive aquatic swarm robotics system," in *OCEANS 2016 - Shanghai*, 10-13 April 2016 2016, pp. 1-7.

[31] H. Khodr, U. Ramage, K. Kim, A. G. Ozgur, B. Bruno, and P. Dillenbourg, "Being Part of the Swarm: Experiencing Human-Swarm Interaction with VR and Tangible Robots," presented at the Proceedings of the 2020 ACM Symposium on Spatial User Interaction, Virtual Event, Canada, 2020.

[32] R. Eberhart and J. Kennedy, "A new optimizer using particle swarm theory," in *MHS'95. Proceedings of the Sixth International Symposium on Micro Machine and Human Science*, 4-6 Oct. 1995 1995, pp. 39-43.

[33] "Coppelia Robotics." https://www.coppeliarobotics.com/ (accessed September 26, 2024).

[34] "Qualtrics." https://research.unsw.edu.au/qualtrics (accessed March 25, 2025).

[35] M. R. Endsley, "Situation awareness global assessment technique (SAGAT)," in *Proceedings of the IEEE 1988 National Aerospace and Electronics Conference*, 23-27 May 1988 1988, pp. 789-795 vol.3.

[36] R. M. Taylor, "Situational awareness rating technique (SART): The development of a tool for aircrew systems design," in *Situational awareness*: Routledge, 2017, pp. 111-128.